\documentstyle[prl,aps,epsf,floats]{revtex}
\input{psfig.tex}
\begin{document}
\def\gtorder{\mathrel{\raise.3ex\hbox{$>$}\mkern-14mu
             \lower0.6ex\hbox{$\sim$}}}
\def\ltorder{\mathrel{\raise.3ex\hbox{$<$}\mkern-14mu
             \lower0.6ex\hbox{$\sim$}}}



\title{Domain wall dominated universes}

\author{Richard A. Battye${}^1$, Martin Bucher${}^1$ and David Spergel${}^2$}
\address{${}^1$ Department of Applied Mathematics and Theoretical
Physics, University of Cambridge, \\ Silver Street, 
Cambridge CB3 9EW, UK\\
${}^2$Department of Astrophysical Sciences, Princeton University,
Princeton, NJ 08544, USA}

\maketitle
\begin{abstract}
We consider a cosmogony with a dark matter component consisting of  a
network of frustrated domain walls. Such a network  provides a solid
dark matter component with $p=-(2/3)\rho $ that remains unclustered on
small scales and with $\Omega_{\rm dw}\approx 0.7$ can reconcile a
spatially  flat universe with the many observations indicating
$\Omega_{\rm m}\approx 0.3.$  Because of its large negative pressure,
this component can explain the recent observations indicating an
accelerating  universe without recourse to a non-vanishing cosmological
constant.  We explore the viability of this proposal and prospects for
distinguishing it from other kinds of proposed dark matter with
significant negative pressure. 
\end{abstract}


Recent observations of apparent luminosities of Type Ia supernovae
(SNIa) at moderate redshift $(\langle z\rangle\approx 0.6)$ suggest
that the expansion of the universe is  now accelerating, indicating a
form of matter with a significant negative pressure~\cite{SN}. From a
theoretical standpoint, these observations are remarkable because most
conceivable contributions to the stress-energy entail a  positive
(radiation) or vanishing (matter) pressure, rather than a large
negative pressure. 

Two proposals to explain these observations  are a non-vanishing
cosmological  constant $(\Lambda)$ or a very slowly rolling scalar
field, often dubbed {\it quintessence}. Both proposals, however, are
plagued with  formidable fine tuning problems. The fine tuning
problems of $\Lambda\ne 0$ are well known~\cite{cosmo}. For the
alternative of a slowly rolling scalar field~\cite{quint}
extraordinarily flat potentials are required,
so that the field is unable to roll to the true minimum by the present day.
If the necessary flatness of such a potential is characterized by a
mass scale $m$, one requires  $m\sim 10^{-33}~{\rm eV}$. 

In this {\it letter} we suggest another form of dark matter with
significant negative pressure---a solid dark matter (SDM)
component with the properties of a  relativistic solid~\cite{bs}. 
The term {\it solid} here 
denotes a substance with a harmonic, non-dissipative
resistance to pure shear (volume preserving) deformations. 
A perfect fluid with negative pressure
($w=p/\rho <0$) would have an imaginary sound speed, indicating
instabilities most severe on the  smallest scales. For a solid,
however, with a sufficiently  large shear modulus these instabilities
are removed. Since the sound speed of perturbations of the solid
should comprise a substantial fraction of
the speed of light, its Jeans length today is
comparable to the size
of the current horizon, and hence an SDM component  would
remain unclustered on the smaller scales over which $\Omega_{\rm
m}$ is measured and thus evade detection.

To compute the effect of an exotic dark matter component on the
evolution of cosmological perturbations, more is required than just
knowing how $w$ evolves throughout  cosmic history~\cite{hu}. As we
will see, it is possible to construct dark matter components for which
$w$ agrees at all times, but which respond differently to cosmological
perturbations.  Because of the non-vanishing shear modulus in the SDM
case, large anisotropic stresses are generated, whereas in models with
a $\Lambda\ne 0$ or a slowly rolling scalar field these  stresses
vanish.  Furthermore, long-wavelength  gravity waves entering the horizon
at late times acquire an effective  mass because of the energetic cost
associated with pure shear deformations of spacetime.  In a previous
paper~\cite{bs} two of us (MB and DNS) developed a formalism 
for computing the evolution
of cosmological perturbations in the presence of a SDM
component and computed 
the cosmic microwave background (CMB) anisotropy on 
large scales. Here we discuss the
viability of such scenarios for creating cosmic structure and the
prospects for distinguishing SDM from other types of dark matter with
significant negative pressure using future measurements of the CMB. A
subsequent paper~\cite{BBSb} will elucidate in more detail the
differences between these scenarios and other models, such as
quintessence and $\Lambda\ne 0.$

A possible microphysical origin for SDM is a frustrated network of
topological defects, either cosmic strings or domain walls (see
ref.~\cite{VS} for a review of cosmic defects). 
The standard picture for the formation
and evolution of topological defects is that of a random network of
defects produced in a cosmological phase transition which 
then evolves toward a self-similar scaling regime with the number
of defects within a Hubble volume approaching a fixed number~\cite{VV}.
However, such behavior relies on the ability of the defects to
untangle and lose energy essentially as fast as allowed by causality,
and in more complicated models this need not occur.
In particular, in theories with several species of defects
or with non-Abelian symmetries, topological obstructions to such
untangling may arise~\cite{tpobs} and defect-dominated evolution can
occur. The basic features of such evolution would be that
after the decay of an initial transient the number of defects per co-moving
volume approaches a constant\cite{V,PSM}. For the simple domain walls of
most interest here this implies an equation 
of state with $w=-2/3$ ($\rho_{\rm dw}\sim a^{-1}$) and for strings $w=-1/3$ 
($\rho_{\rm str}\sim a^{-2}$), although 
other values of $w$ may be possible as well. Simulations of non-Abelian cosmic
strings~\cite{PSM} indicate behavior leading to a string-dominated
universe and simulations of domain walls~\cite{kubo} find that the
density of walls in a model with several vacua falls less rapidly than would be required by scaling. In axion
models, where an additional symmetry breaking $U(1)\to Z_N$ allows for
$N$ domain walls to meet at a string, scaling behavior has been
observed~\cite{rps}. The precise criterion for when defect
domination takes place is not clear at present, although it seems
clear that such models can exist. This is currently under
investigation.

The symmetry breaking scale $\eta$ required for the  formation of a
network of frustrated domain walls  with $\Omega _{\rm dw}\sim 1$
today is around $\eta\approx 100~{\rm KeV}$, assuming a phase
transition at $T\approx\eta$, with initially one wall per
horizon volume and a network immediately  settling down
to an equilibrium configuration which is subsequently swept along by the
Hubble flow. This estimate, however, is subject to  considerable
uncertainties in  either direction. For example, a long transient
before settling down to an equilibrium  configuration could
considerably raise $\eta$.  The mean separation between walls
is approximately 30 parsecs, much smaller than the scale on which the
network  clusters in response to gravitational
perturbations.  It is therefore justified to treat the network as a
continuum solid when studying its response to cosmological
perturbations. 

In a recent article~\cite{WCOS} it was suggested that a quintessence
model with a flat geometry $w\approx -2/3$, $\Omega_{\rm c}=0.25$, $\Omega_{\rm
b}=0.05$, $h=0.65$, $n=1$ gave a adequate fit to data probing a range of
epochs and scales, where
$\Omega_{\rm X}$ is the fractional density relative to the critical
density in particle species X (X=c
is CDM, X=b is baryons), the Hubble constant is $H_0=100h~{\rm km}\,{\rm
sec}^{-1}\,{\rm Mpc}^{-1}$ and $n$ is the spectral index of the
initial density perturbations. One might, therefore, wonder whether a
domain wall dominated model with $\Omega_{\rm dw}\approx 0.7$ might
fit the data equally well and, in fact, a number of the 
calculations presented in ref.~\cite{WCOS} apply equally well to the
case of a SDM model. However, important differences in how
perturbations evolve once the SDM component comes to
dominate must be considered for an
accurate comparison to the data. Due to the apparently
good fit of the $w=-2/3$ case, most of the discussion focuses on
the domain wall dominated case, but we also discuss the
string dominated case and other values of $w$. Except where
expressly stated, all our results use the parameter choices listed above.

\begin{figure}
\centerline{\psfig{file=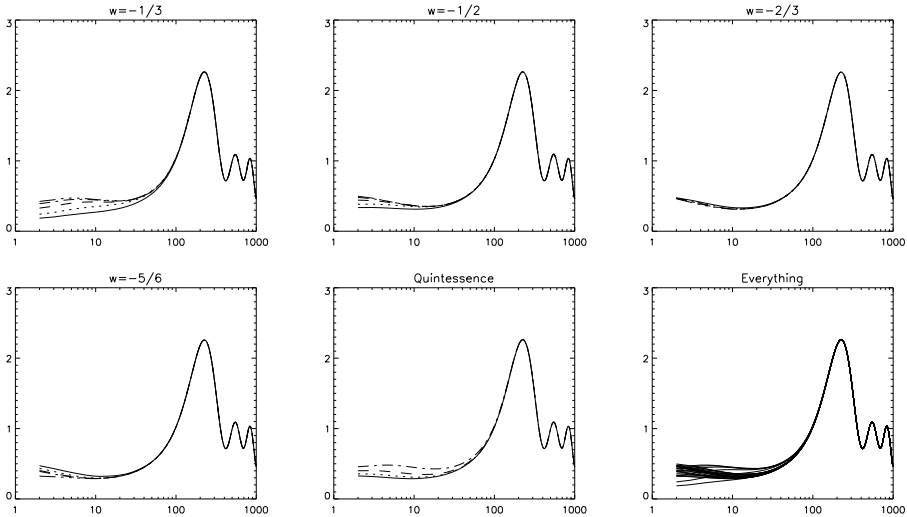,width=5.0in}}
\caption{{\bf Rescaled CMB Moments for SDM Models.}
In the first four panels $\ell (\ell +1)C_\ell $ versus $\ell $
is plotted for $w=-1/3, -1/2, -2/3, -5/6$ for
several values of the longitudinal sound speed for a 
universe with $\Omega_{\rm m}=0.3, \Omega_{\rm sdm}=0.7.$
The solid curve corresponds to $c_s=0.0$ and $c_s=0.2, 0.4,
0.6, 0.8$ are indicated by the line styles
dotted, dashed, dot-dash, and
dot-dot-dash, respectively.
rescaled so that the horizon size at recombination
subtends the same angle in the sky today as for a $w=-1$
model with the same $\Omega _{\rm m}.$
The bottom middle panel indicates the
rescaled and renormalized moments for the
quintessence models with $w=-1/3, -1/2, -2/3, -5/6,$
and the bottom right panel shows all models
of the previous panels for comparison.}
\label{cmb}
\end{figure}

We have included the evolution of a SDM component and its effects on the
other perturbation variables into the standard Einstein-Boltzmann
solver CMBFAST. As well as specifying $w$, this requires the
introduction of a another parameter $c_s$ which is the sound speed of
scalar perturbations in the solid, related to the the vector sound
speed, $c_v$, by $c_s^2=4c_v^2/3+w$. For a given $w$, the evolution of
the Newtonian potentials  $\Phi$ and $\Psi$ at late times differs 
as one varies $c_s$, and also differs from quintessence
models. This leads to distinct integrated Sachs-Wolfe (ISW) contributions
to the CMB anisotropies. At
recombination, when the density of the SDM component is  negligible
(for $w\ltorder -1/3$), these differences are suppressed and  the
small-angle anisotropies are identical for the same 
initial fluctuations. For this reason, when the results of our
computations are plotted in Fig.~\ref{cmb}, they are normalized at an
angular scale corresponding to $\ell=500$ and spectra for different
values of $c_s$ are then identical for $\ell>100$.
Models with different $w$, however, have different peak
positions with all the other cosmological parameters fixed, 
since the angular diameter distance $\ell_D=k_{\rm p}\eta_0$ depends
on $w.$
The wavenumber $k_{\rm p}$ corresponds to the first acoustic
peak~\cite{white} and any change in $\eta_0$ due to a variation in $w$ can
be offset by modifying $k_{\rm p}$, as discussed in
ref.~\cite{HWDCS}. In Fig.~\ref{cmb}, the angular scale has been
rescaled so that all the models have the angular diameter distance of
the model with $w=-1$. Distinguishing among the range of
models considered will require accurate 
measurements of the CMB anisotropies on large scales. 

\begin{table}
\centering
\begin{tabular}{||c|c||c|c|c|c|c|c|c|c|c||}
\hline
\multicolumn{2}{||c||}{}& \multicolumn{4}{c|}{$w=-1/3$}
&\multicolumn{4}{c|}{$w=-2/3$}&$w=-1$\\ \cline{3-11}
\multicolumn{2}{||c||}{}&$c_s=0.0$ &$c_s=0.2$ &$c_s=0.6$ & Q
&$c_s=0.0$ &$c_s=0.2$ &$c_s=0.6$ & Q &$\Lambda $\\ \hline\hline
&$c_s=0.0$
&      0.0&      9.6&      26.1&      26.6&      5.9&
      4.7&      4.1&      2.6&      69.8\\
\cline{2-11}
&$c_s=0.2$
&      10.9&      0.0&      5.2&      5.8&      5.5&
      7.0&      7.4&      7.1&      33.1\\
\cline{2-11}
&$c_s=0.6$
&      34.1&      6.2&      0.0&     0.12&      14.5&
      18.1&      19.6&      21.0&      20.9\\
\cline{2-11}
\raisebox{4ex}[0pt]{$w=-1/3$}
&Q       
&      35.6&      7.2&     0.13&      0.0&      13.9&
      17.5&      19.1&      20.9&      22.1\\
\hline
&$c_s=0.0$
&      8.3&      5.7&      12.7&      12.1&      0.0&
     0.18&     0.38&     0.93&      53.9\\
\cline{2-11}
&$c_s=0.2$
&      6.8&      7.0&      15.6&      15.0&     0.18&
      0.0&    0.05&     0.43&      58.6\\
\cline{2-11}
&$c_s=0.6$
&      5.9&      7.2&      16.7&      16.2&     0.37&
    0.05&      0.0&     0.25&      59.9\\
\cline{2-11}
\raisebox{4ex}[0pt]{$w=-2/3$}
&Q
&      3.4&      6.6&      17.8&      17.5&     0.86&
     0.39&     0.22&      0.0&      61.3\\
\hline 
$w=-1$&$\Lambda $
&      93.6&      40.5&      24.9&      26.5&      69.7&
      76.4&      78.3&      80.1&      0.0\\
\hline
\end{tabular}

\vskip 10pt
\caption{The distinguishability of various types of solid dark matter
and quintessence using the CMB.
The table above indicates the expectation value of the
natural logarithm of the likelihood of the model
corresponding to the column relative to that of the
row assuming that the CMB moments were generated
by the model corresponding to the column.
For all models $\Omega _m=0.3$. 
As described in the text, the CMB multipole moments
of the models were rescaled so that the overall amplitudes
and the shapes coincide at large $\ell .$ These relative
likelihoods take into account only cosmic variance and
assume full sky coverage. Observational errors and the fact
that varying other cosmological parameters affects the
shape of the low-$\ell $ moments
is not taken into account.
}\label{tab}
\end{table}

Table 1 examines the significance of these differences by
comparing flat models with identical cosmological parameters, but with
differing properties for the component with significant negative
pressure rescaled and normalized as described above.  An estimate of
the significance may be obtained by ignoring
observational uncertainties and incomplete sky  coverage because of
the galaxy and considering only the effect of cosmic  variance,
arising from the fact that for each $\ell $ one is able to observe
only $(2\ell +1)$ realizations of a Gaussian random process. Assuming
that a model A is correct, the expectation value of the natural
logarithm of the relative likelihood of model A relative to a model B
is 
\begin{equation}
\left< {\rm ln}  \left( {P(\{ a_{\ell m}\} | A )\over P(\{ a_{\ell
m}\} | B )} \right) \right>_A 
=-{1\over 2}\sum _\ell (2\ell +1)\left[ 1-{C_\ell ^{(A)}\over C_\ell
^{(B)}}+ {\rm ln}\left({C_\ell ^{(A)}\over C_\ell ^{(B)}}\right)
\right] .
\end{equation}
For a domain wall dominated universe $(w=-2/3)$ we find it
virtually impossible to distinguish between  differing sound speeds or
between SDM and  quintessence, while for  $w=-1/3$ we
observe substantial differences between  differing sound speeds, and
between SDM and  quintessence.  Moreover, all the models are
distinguishable from $\Lambda $ model with the same value of
$\Omega_{\rm m}$. The relative likelihood 
of different models increases as $\Omega_{\rm m}$ 
is decreased~\cite{BBSb}. 

\begin{figure}
\setlength{\unitlength}{1cm}
\begin{minipage}{8.0cm}
\leftline{\psfig{file=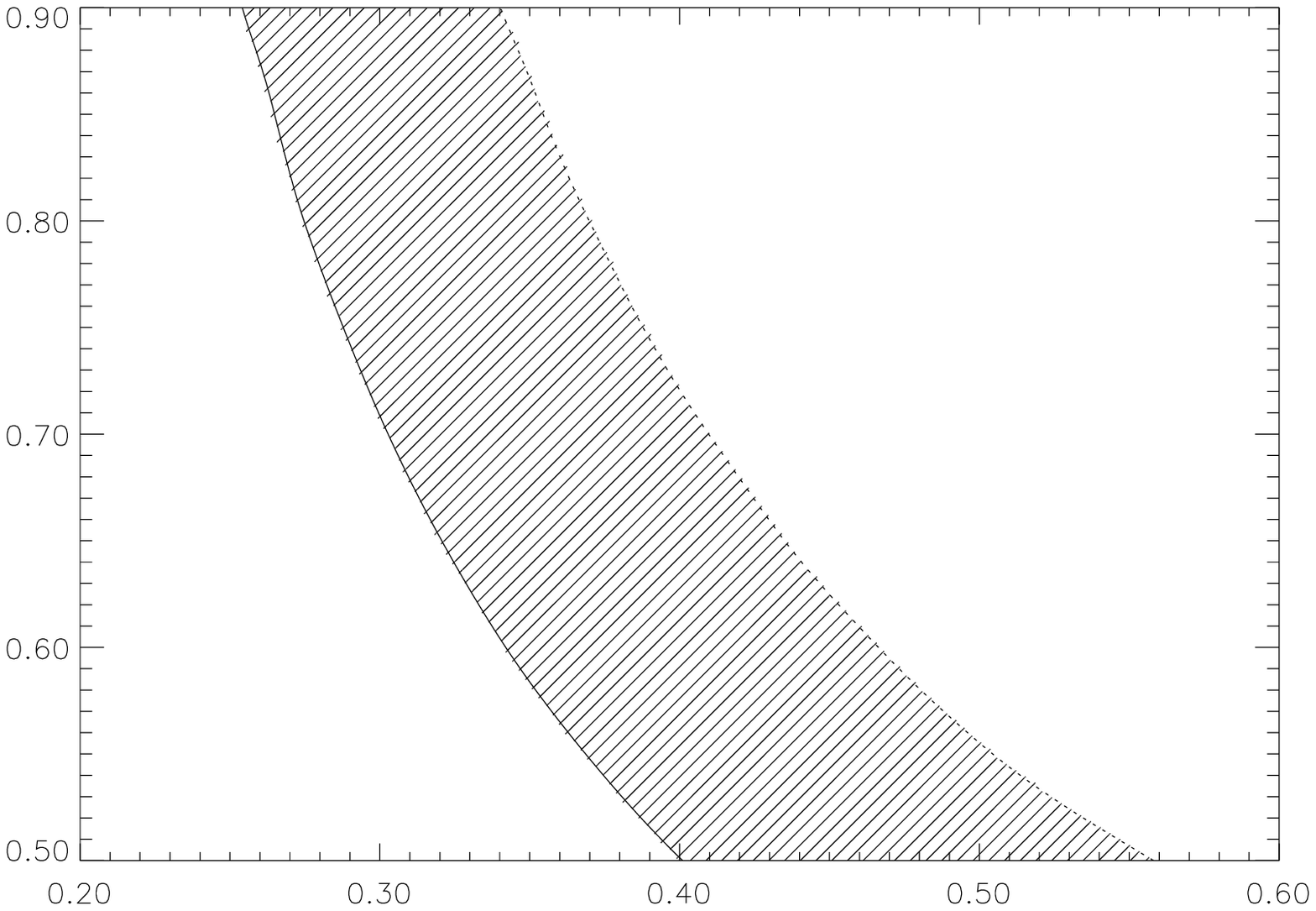,width=3.4in}}
\end{minipage}\hfill
\begin{minipage}{8.0cm}
\rightline{\psfig{file=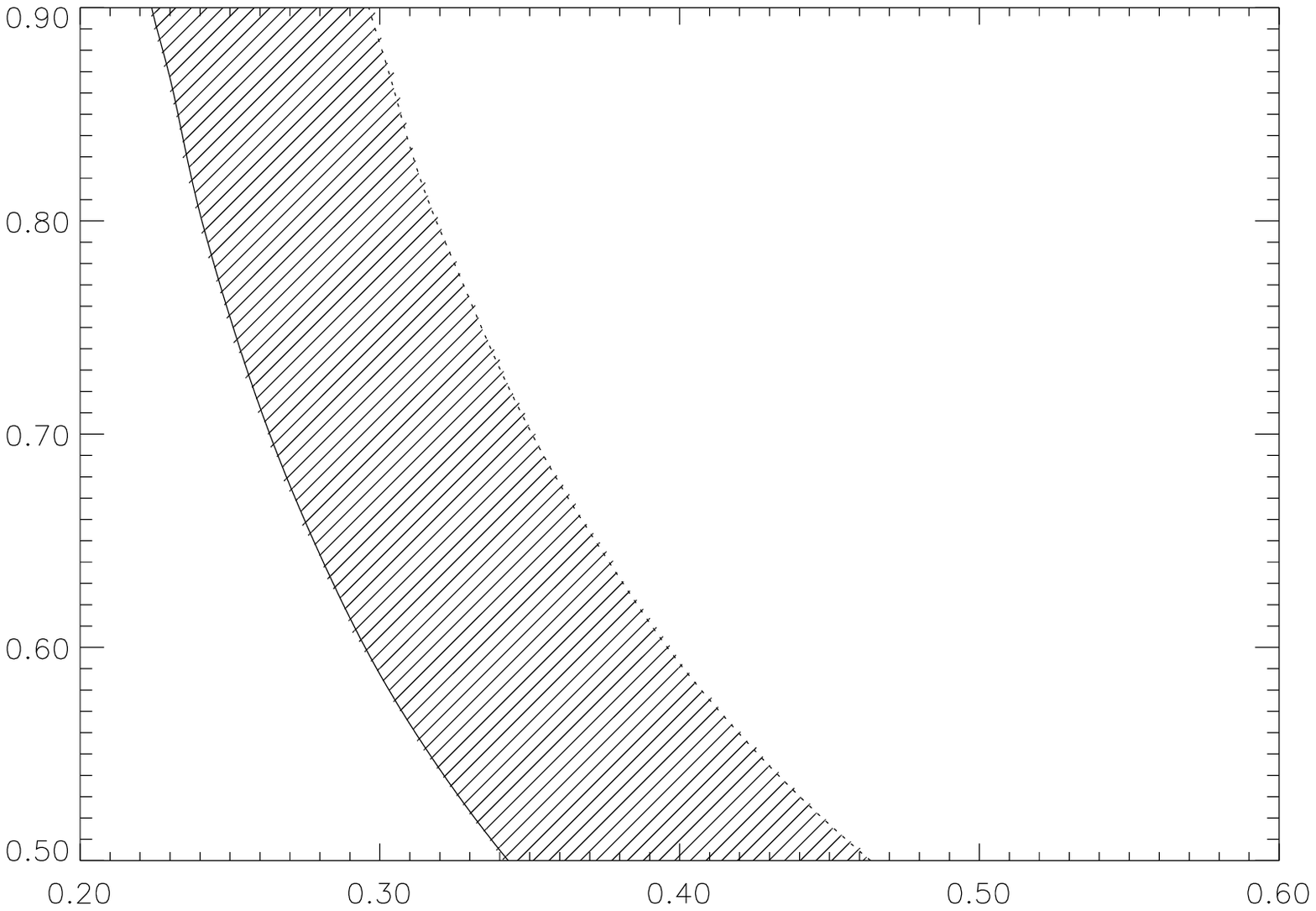,width=3.4in}}
\end{minipage}
\caption{Domain wall dominated models 
compatible with constraints on $\sigma_8$ from X-ray 
clusters. The solid line indicates
the stringent lower limit and 
the dotted line the upper limit, which can be weakened by 
including a tensor component. The vertical and horizontal
axes are $h$ and $\Omega_{\rm m},$ respectively. The left 
and right panels indicate the spectral indices 
$n=1.0$ and $n=1.1$, respectively. In both cases the 
longitudinal sound speed is $c_s=0.2.$}
\label{lss}
\end{figure}

An important test for any cosmological model is whether it can create the large-scale 
structure (LSS) seen today when normalized to the 
the CMB anisotropies observed by COBE.
The simplest version of this 
is to compare the COBE normalized value of $\sigma_8$, 
the variance of the density 
field in spheres of radius $8h^{-1}{\rm Mpc},$ with that obtained from 
the observations of X-ray clusters~\cite{clusters}. 
Models with $w=-1$ pass this test, 
but two effects can cause models with $w>-1$ to predict a lower 
$\sigma _8$ when normalized to COBE:
a large integrated Sachs-Wolfe component reduces the predicted 
value of $\sigma_8$, and for larger $w$ 
the growth of perturbations becomes stunted earlier because 
the SDM (or quintessence) component begins to dominate the universe
earlier. 
To investigate the viability of 
these models we compute the COBE normalized $\sigma_8$ for $w=-2/3$ 
and a range of values in the $\Omega_{\rm m}-h$ plane, which are then  compared to 
the recently computed observational value~\cite{clusters} 
$\sigma_8\Omega_{\rm m}^{\gamma}=(0.5-0.1\Theta)\pm 0.1$, 
where $\gamma=0.21-0.22w+0.33\Omega_{\rm m}+0.25\Theta$ 
and $\Theta=(n-1)+(h-0.65)$, valid in the 
range of parameters considered. Fig.~\ref{lss} shows the regions 
of $\Omega_{\rm m}-h$ plane which pass this test for spectral 
indices $n=1.0$ and $n=1.1.$ 
For $n=1.0,$ the preferred values for quintessence models $h=0.65$ and 
$\Omega_{\rm m}=0.3$ are marginally incompatible with the quoted values for 
$\sigma_8$, but this situation is easily rectified by increasing 
$h$. Furthermore, these parameters are compatible with $n=1.1$, 
although the viability of scenarios with $w=-1/3$ would require 
even larger values of $n$. The constraint 
from the upper bound on $\sigma_8$ is weaker 
because of the possibility of including
a tensor contribution to the CMB anisotropies~\cite{BBSb}.

We have shown that the SDM proposal, and in 
particular a domain wall dominated universe, is compatible with LSS via 
the COBE normalization/$\sigma_8$ test and that the inclusion of 
a SDM component, as opposed to quintessence, can have some 
potentially interesting effects on the CMB anisotropies.
If the SNIa results are confirmed, an 
important question will be how to distinguish 
between SDM and quintessence models.
Some possibilities include 
cross-correlating the CMB with LSS using, for example the X-ray 
background~\cite{turok}, or the gravitational lensing of the 
CMB~\cite{silk}. Observations of gravitational lensing
can measure the evolution of the power spectrum with
redshift~\cite{hutwo,tyson}, and in particular the 
proposed Dark Matter Telescope would measure 
the growth of perturbations  
with high accuracy. 

We would like to thank Brandon Carter, Neil Turok, and Alexander
Vilenkin for useful discussions, and Uros Seljak and Matias
Zaldariagga for the use of CMBFAST. RAB was funded by Trinity College,
MB was supported by PPARC, and DNS was supported in part by  MAP. 

\def\jnl#1#2#3#4#5#6{\hang{#1, {\it #4\/} {\bf #5}, #6 (#2).} }
\def\jnltwo#1#2#3#4#5#6#7#8{\hang{#1, {\it #4\/} {\bf #5}, #6; {\it
ibid} {\bf #7} #8 (#2).} }  \def\prep#1#2#3#4{\hang{#1, #4.} }
\def\proc#1#2#3#4#5#6{{#1 [#2], in {\it #4\/}, #5, eds.\ (#6).} }
\def\book#1#2#3#4{\hang{#1, {\it #3\/} (#4, #2).} }
\def\jnlerr#1#2#3#4#5#6#7#8{\hang{#1 [#2], {\it #4\/} {\bf #5}, #6.
{Erratum:} {\it #4\/} {\bf #7}, #8.} } \def\prl{Phys.\ Rev.\ Lett.}
\def\pr{Phys.\ Rev.}  \def\pl{Phys.\ Lett.}  \def\np{Nucl.\ Phys.}
\def\prp{Phys.\ Rep.}  \def\rmp{Rev.\ Mod.\ Phys.}  \def\cmp{Comm.\
Math.\ Phys.}  \def\mpl{Mod.\ Phys.\ Lett.}  \def\apj{Ap.\ J.}
\def\apjl{Ap.\ J.\ Lett.}  \def\aap{Astron.\ Ap.}  \def\cqg{Class.\
Quant.\ Grav.}  \def\grg{Gen.\ Rel.\ Grav.}  \def\mn{MNRAS}
\def\ptp{Prog.\ Theor.\ Phys.}  \def\jetp{Sov.\ Phys.\ JETP}
\def\jetpl{JETP Lett.}  \def\jmp{J.\ Math.\ Phys.}  \def\zpc{Z.\
Phys.\ C} \def\cupress{Cambridge University Press} \def\pup{Princeton
University Press} \def\wss{World Scientific, Singapore}
\def\oup{Oxford University Press}

\end{document}